\documentclass{article}

\usepackage{amsmath}
\usepackage{amssymb}
\usepackage{amsthm}
\usepackage{amscd}
\usepackage[all]{xy}

\newtheorem{thm}{Theorem}[section]
\newtheorem{prop}[thm]{Proposition}
\newtheorem{lem}[thm]{Lemma}

\newtheorem{ithm}{Theorem}

\theoremstyle{definition}
\newtheorem{dfn}[thm]{Definition}

\theoremstyle{remark}
\newtheorem{rem}{Remark}
\newtheorem*{acknowledgment}{Acknowledgment}

\newcommand{\C}{\mathbb{C}}
\newcommand{\N}{\mathbb{N}}
\newcommand{\R}{\mathbb{R}}
\newcommand{\T}{\mathbb{T}}
\newcommand{\Z}{\mathbb{Z}}

\newcommand{\Aa}{\mathfrak{A}}
\newcommand{\U}{\mathfrak{U}}

\newcommand{\G}{\mathcal{G}}

\newcommand{\Cech}{$\check{\textrm{C}}$ech {}}
\newcommand{\Poincare}{Poincar\'e {}}
\newcommand{\dvol}{d\mathrm{vol}}
\newcommand{\im}{i}
\renewcommand{\d}{\partial}

\newcommand{\un}{\underline}

\def\h#1{ \widehat{#1} }
\def\til#1{ \tilde{#1} }
\def\SDC#1{ \Z({#1})^\infty_D }
\def\CSDC#1{ \Z({#1})^\infty_{D, \C}}

\title{Central extensions of gauge transformation groups of 
higher abelian gerbes}

\author{Kiyonori Gomi
\thanks{The author's research is supported by Research Fellowship of the Japan Society for the Promotion of Science for Young Scientists.}}

\date{}

\begin{document}

\maketitle

\begin{abstract}
We construct a central extension of the smooth Deligne cohomology group of a compact oriented odd dimensional smooth manifold, generalizing that of the loop group of the circle. While the central extension turns out to be trivial for a manifold of dimension 3, 7, 11, \ldots, it is non-trivial for 1, 5, 9, \ldots. In the case where the central extension is non-trivial, we show an analogue of the Segal-Witten reciprocity law. 
\end{abstract}


\section{Introduction}
\label{sec:introduction}

The \textit{smooth Deligne cohomology} \cite{Bry,D-F,E-V} of a smooth manifold $X$ is defined to be the hypercohomology $H^q(X, \SDC{p})$  of the complex of sheaves
$$
\SDC{p} : 
\Z {\longrightarrow}
\un{A}^0 \overset{d}{\longrightarrow}
\un{A}^1 \overset{d}{\longrightarrow}
\cdots \overset{d}{\longrightarrow}
\un{A}^{p-1} \longrightarrow
0 \longrightarrow \cdots,
$$
where $\Z$ is the constant sheaf, located at degree $0$, and $\un{A}^q$ is the sheaf of germs of $\R$-valued differential $q$-forms. If $X$ is compact, then we can identify the abelian group $H^p(X, \SDC{p})$ with $(A^{p-1}(X)/A^{p-1}(X)_{\Z}) \times H^{p}(X, \Z)$ by a non-canonical isomorphism, where $A^{p-1}(X)$ is the group of $(p-1)$-forms on $X$, and $A^{p-1}(X)_\Z$ is the subgroup consisting of closed integral $(p-1)$-forms.

\medskip

Heuristically, we can interpret $H^{n+2}(X, \SDC{n+2})$ as the group classifying ``abelian $n$-gerbes with connection on $X$.'' Besides the interpretation, we can also interpret $H^{n+1}(X, \SDC{n+1})$ as the ``gauge transformation group of an abelian $n$-gerbe on $X$.'' For example, if $n = 0$, then $H^1(X, \SDC{1})$ is naturally isomorphic to the group $C^\infty(X, \T)$ of smooth functions on $X$ with values in the unit circle $\T = \{ u \in \C |\ |u| = 1 \}$. The group is nothing but the gauge transformation group of a principal $\T$-bundle ($0$-gerbe) over $X$. 

Gauge transformation groups of principal bundles are fundamental ingredients not only in mathematics but also in physics. Hence it would be meaningful to study smooth Deligne cohomology groups as gauge transformation groups of higher abelian gerbes.

\medskip

On the circle $S^1$, the gauge transformation group of a principal bundle with connected structure group is often identified with a loop group.  As is well-known \cite{P-S}, the loop group of a compact Lie group has non-trivial central extensions by $\T$. Let $\h{L\T}$ denote the \textit{universal central extension} of the loop group $L\T = C^\infty(S^1, \T)$. In \cite{Bry-M2}, Brylinski and McLaughlin provided geometric constructions of a central extension of $L\T = H^1(S^1, \SDC{1})$ which is isomorphic to $\h{L\T}/\Z_2$. We can describe one of their constructions, using only basic tools of smooth Deligne cohomology. As a straight generalization of the description, we can construct a central extension of $H^{n+1}(M, \SDC{n+1})$ by $\T$ for any compact oriented smooth $(2n + 1)$-dimensional manifold $M$, which is the subject of this paper. 

\medskip

We explain the construction more precisely. The basic tools we use are the \textit{cup product} and the \textit{integration} for smooth Deligne cohomology. Let $X$ be a smooth manifold. The cup product is a natural homomorphism
$$
\cup : \
H^p(X, \SDC{p}) \otimes_{\Z} H^q(X, \SDC{q}) \longrightarrow 
H^{p+q}(X, \SDC{p+q})
$$
which is associative and (graded) commutative \cite{Bry,E-V}. For the integration to be considered, we suppose that $X$ is a compact oriented smooth $d$-dimensional manifold without boundary. Then the integration is a natural homomorphism
$$
\int_X : \ H^{d+1}(X, \SDC{d+1}) \longrightarrow \R/\Z.
$$
By using \Cech cohomology, we can describe the operations above explicitly. 

Let $n$ be a non-negative integer fixed. We put $\G(X) = H^{n+1}(X, \SDC{n+1})$ for a smooth manifold $X$. For a compact oriented smooth $(2n+1)$-dimensional manifold $M$ without boundary, a group 2-cocycle $S_M : \G(M) \times \G(M) \to \R/\Z$ is given by $S_M(f, g) = \int_{M} f \cup g$. Now we define a central extension $\til{\G}(M)$ to be the set $\G(M) \times \T$ endowed with the group multiplication
$$
(f, u) \cdot (g, v) = (f + g, uv \exp2\pi\im S_M(f, g)).
$$

In general, we can make $\G(M)$ and $\til{\G}(M)$ into infinite dimensional Lie groups. However, to avoid discussion not essential to our aim, we will not treat such Lie group structures in this paper.

\medskip

One result of this paper is the (non-)triviality of $\til{\G}(M)$ as a central extension of $\G(M)$. 

\begin{ithm} \label{ithm:central_ext}
Let $n$ be a non-negative integer, and $M$ a compact oriented smooth $(2n+1)$-dimensional manifold without boundary. 

(a) If $n = 2k$, then $\til{\G}(M)$ is non-trivial.

(b) If $n = 2k+1$, then $\til{\G}(M)$ is trivial.
\end{ithm}

As a consequence, we obtain non-trivial central extensions in the case that the dimension of $M$ is 1, 5, 9, \ldots. In particular, when $n = 0$ and $M = S^1$, we recover the central extension $\h{L\T}/\Z_2$.

\medskip

The other result of this paper is an analogue of the \textit{Segal-Witten reciprocity law} \cite{Bry-M1,Bry-M2}, which is a basic property of central extensions of loop groups. To state our result, we introduce the following complex of sheaves on a smooth manifold $X$:
$$
\CSDC{p} :\ 
\Z \longrightarrow
\un{A}^0_\C \overset{d}{\longrightarrow}
\un{A}^1_\C \overset{d}{\longrightarrow}
\cdots \overset{d}{\longrightarrow}
\un{A}^{p-1}_\C \longrightarrow
0 \longrightarrow \cdots,
$$
where $\un{A}^q_\C$ is the sheaf of germs of $\C$-valued differential $q$-forms. We can think of the hypercohomology $H^q(X, \CSDC{p})$ as a complexification of $H^q(X, \SDC{p})$. When a positive integer $n$ is fixed, we put $\G(X)_\C = H^{n+1}(X, \CSDC{n+1})$ for a smooth manifold $X$.

Suppose that $n$ is even. We put $n = 2k$. If $M$ is a compact oriented smooth $(4k+1)$-dimensional manifold without boundary, then a construction similar to that of $\til{\G}(M)$ provides a non-trivial central extension $\til{\G}(M)_\C$ of $\G(M)_{\C}$ by $\C^* = \C - \{ 0 \}$. Let $W$ be a compact oriented smooth $(4k+2)$-dimensional manifold with boundary. Then we obtain a homomorphism $r : \G(W)_\C \to \G(\d W)_\C$ by the restriction, and a central extension $r^*\til{\G}(\d W)_\C$ of $\G(W)_\C$ by the pull-back.

Notice here that there exists an epimorphism $\delta : \G(W)_\C \to A^{2k+1}(W, \C)_\Z$ in general, where $A^{2k+1}(W, \C)_\Z$ is the group of $\C$-valued closed $(2k+1)$-forms on $W$ with integral periods. Suppose that $W$ is equipped with a Riemannian metric. Because the Hodge star operator on $(2k+1)$-forms satisfies $** = -1$, we can introduce a subgroup $\G(W)^+_\C$ in $\G(W)_\C$ by
$$ 
\G(W)^+_\C = 
\{ f \in \G(W)_\C |\ 
\delta(f) - \im * \delta(f) = 0\}.
$$
The generalization of the Segal-Witten reciprocity law in this paper is:

\begin{ithm} \label{ithm:Segal_Witten}
Let $k$ be a non-negative integer, and $W$ a compact oriented smooth $(4k+2)$-dimensional Riemannian manifold with boundary. Then $r^*\til{\G}(\d W)_\C$ splits over the subgroup $\G(W)^+_\C$. 
$$
\xymatrix{
1 \ar[r] & 
\C^* \ar[r] & 
r^*\til{\G}(\d W)_\C \ar[r] & 
\G(W)_\C \ar[r] & 
1 \\
       &
       &
       & 
\G(W)^+_\C \ar@{-->}[lu]  \ar@{^{(}->}[u] &
}
$$
\end{ithm} 

In the case of $k = 0$, a Riemannian metric on $W$ makes $W$ into a Riemann surface, and $\G(W)^+_\C$ is isomorphic to the group of holomorphic $\C^*$-valued functions on $W$. Hence Theorem \ref{ithm:Segal_Witten} recovers the usual Segal-Witten reciprocity law for $\h{L\T}/\Z_2$ in this case.

\bigskip

We remark that the Segal-Witten reciprocity law and \textit{positive energy representations} \cite{P-S} of a central extension of a loop group are used in a (formal) definition of the space of conformal blocks in Wess-Zumino-Witten model. This motivates us to construct infinite dimensional representations of $\til{\G}(M)$ for $n = 2k > 0$, and a construction will be given in a subsequent paper \cite{Go}.

\bigskip

The present paper is organized as follows. In Section \ref{sec:Deligne}, we review smooth Deligne cohomology. We explicitly give the cup product and the integration. Though unnecessary to prove the results of this paper, geometric interpretations of smooth Deligne cohomology are formally explained in the section. In Section \ref{sec:central_ext}, we construct the central extension $\til{\G}(M)$, and prove the (non-)triviality (Theorem \ref{thm:non_trivial} and Theorem \ref{thm:trivial}). Finally, in Section \ref{sec:reciprocity_law}, we prove the generalization of the Segal-Witten reciprocity law.


\section{Smooth Deligne cohomology}
\label{sec:Deligne}

We recall here the definition of smooth Deligne cohomology \cite{Bry,D-F,E-V}. The cup product and the integration are introduced by using \Cech cohomology. We also explain geometric interpretations of smooth Deligne cohomology groups in a formal fashion.


\subsection{Definition of smooth Deligne cohomology}

Let $X$ be a smooth manifold. For a non-negative integer $q$, we denote by $\un{A}^q$ the sheaf of germs of $\R$-valued differential $q$-forms on $X$.

\begin{dfn}
For a non-negative integer $p$, we define the \textit{smooth Deligne complex} $\SDC{p}$ to be the following complex of sheaves:
$$
\SDC{p} : 
\Z \longrightarrow
\un{A}^0 \overset{d}{\longrightarrow}
\un{A}^1 \overset{d}{\longrightarrow}
\cdots \overset{d}{\longrightarrow}
\un{A}^{p-1} \longrightarrow
0 \longrightarrow \cdots,
$$
where $\Z$ is the constant sheaf and is located at degree $0$ in the complex. The \textit{smooth Deligne cohomology group} $H^q(X, \SDC{p})$ of a smooth manifold $X$ is defined to be the hypercohomology of $\SDC{p}$.
\end{dfn}

\begin{rem}
There are several definitions of smooth Deligne cohomology. For instance, one can use the following complex of sheaves: 
$$
\un{\T} \overset{\frac{1}{2\pi\im}d\log}{\longrightarrow} 
\un{A}^1 \overset{d}{\longrightarrow} 
\cdots \overset{d}{\longrightarrow} 
\un{A}^{p-1} \longrightarrow
0 \longrightarrow \cdots,
$$
where $\un{\T}$ is the sheaf of germs of $\T$-valued smooth functions. The complex of sheaves is quasi-isomorphic to $\SDC{p}$, and its hypercohomology is isomorphic to $H^q(X, \SDC{p})$ under a degree shift.
\end{rem}

\begin{rem}
The smooth Deligne cohomology $H^p(X, \SDC{p})$ is isomorphic to the group of \textit{differential characters} introduced by Cheeger and Simons \cite{Chee-S}. For this fact, we refer the reader to \cite{Bry,E-V}.
\end{rem}

For later convenience, we describe $H^q(X, \SDC{p})$ as a \Cech cohomology. We write $(\SDC{p})^{[j]}$ for the sheaf located at degree $j$ in the complex $\SDC{p}$, and $\til{d} : (\SDC{p})^{[j]} \to (\SDC{p})^{[j+1]}$ for the coboundary operator. When an open cover $\U = \{ U_\alpha \}_{\alpha \in \Aa}$ of $X$ is given, we put
$$
\check{C}^{i, j} = 
\prod_{U_{\alpha_0}, \ldots, U_{\alpha_i}}
\Gamma(U_{\alpha_0 \cdots \alpha_j}, (\SDC{p})^{[j]}),
$$
where $U_{\alpha_0 \cdots \alpha_i} = U_{\alpha_0} \cap \cdots \cap U_{\alpha_i}$. Let $\check{\delta} : \check{C}^{i, j} \to \check{C}^{i+1, j}$ be the \Cech coboundary operator, and $\til{d} : \check{C}^{i, j} \to \check{C}^{i, j+1}$ the coboundary operator induced by that on $\SDC{p}$. From the double complex $(\check{C}^{i, j}, \check{\delta}, \til{d})$, we construct a single complex $(\check{C}^p(\U, \SDC{p}), D)$ by setting $\check{C}^q(\U, \SDC{p}) = \oplus_{q = i + j}\check{C}^{i, j}$. As a convention of this paper, we define the total coboundary operator to be $D = \check{\delta} + (-1)^i\til{d}$ on the component $\check{C}^{i, j}$. 

We denote by $\check{H}^q(\U, \SDC{p})$ the cohomology of the total complex. Then the hypercohomology $H^q(X, \SDC{p})$ is defined to be
$$
H^q(X, \SDC{p}) = 
\varinjlim \check{H}^q(\U, \SDC{p}),
$$
where the direct limit is taken over the ordered set of open covers of $X$. As is well-known, if $\U$ is a \textit{good cover} \cite{B-T,Bry} of $X$, then $H^q(X, \SDC{p})$ is isomorphic to the \Cech cohomology $\check{H}^q(\U, \SDC{p})$.

\medskip

It is easy to see that $H^q(X, \SDC{0})$ is naturally isomorphic to the cohomology with integer coefficients $H^q(X, \Z)$ for all $q$. 

\begin{prop}[\cite{Bry}] \label{prop:Deligne_coh_exact_seq}
Let $p$ be a positive integer.

(a) If $0 \le q < p$, then $H^q(X, \SDC{p}) \cong H^{q-1}(X, \R/\Z)$.

(b) If $p = q$, then $H^p(X, \SDC{p})$ fits into the following exact sequences:
\begin{gather}
0 \to
H^{p-1}(X, \R/\Z) \to
H^p(X, \SDC{p}) \overset{\delta}{\to}
A^p(X)_{\Z} \to 0, 
\label{exact_seq:SDC_forms}
\\
0 \to
A^{p-1}(X) / A^{p-1}(X)_{\Z} \overset{\iota}{\to}
H^p(X, \SDC{p}) \overset{\chi}{\to}
H^p(X, \Z) \to 0,
\label{exact_seq:SDC_cohomology}
\end{gather}
where $A^q(X)$ is the group of $q$-forms on $X$, and $A^q(X)_{\Z} \subset A^q(X)$ is the subgroup of closed integral $q$-forms. 

(c) If $p < q$, then $H^q(X, \SDC{p}) \cong H^q(X, \Z)$.
\end{prop}

\begin{proof}
We can readily show the proposition by using the natural spectral sequences associated with the double complex $\check{C}^{p, q}$, a partition of unity and the \Poincare lemma \cite{B-T,Bry}.
\end{proof}

It would be worth while expressing the homomorphisms $\delta, \iota$ and $\chi$ in Proposition \ref{prop:Deligne_coh_exact_seq} (b) by using \Cech cohomology. We write a $p$-cochain as
$$
(n_{\alpha_0 \cdots \alpha_p}, 
\omega^0_{\alpha_0 \cdots \alpha_{p-1}}, \ldots, 
\omega^{p-1}_{\alpha_0}) \in \check{C}^p(\U, \SDC{p}).
$$
The homomorphisms are expressed as
\begin{align*}
\delta &: \check{H}^p(\U, \SDC{p}) \to A^p(X)_\Z, &
& \delta([n, \omega^0, \ldots, \omega^{p-1}]) = d \omega^{p-1}, \\
\iota &: A^{p-1}(X)/A^{p-1}(X)_\Z \to \check{H}^p(\U, \SDC{p}), &
& \iota(\omega) = [(0, \ldots, 0, \omega|_{U_\alpha})], \\
\chi &: \check{H}^p(\U, \SDC{p}) \to \check{H}^p(\U, \Z) &
& \chi([n, \omega^0, \ldots, \omega^{p-1}]) = [(n)].
\end{align*}

\begin{lem} \label{lem:splitting_Deligne}
If $X$ is compact, then there exists a splitting $\sigma : H^p(X, \Z) \to H^p(X, \SDC{p})$ of the exact sequence (\ref{exact_seq:SDC_cohomology}), so that there is an isomorphism of abelian groups
\begin{equation}
I : \ (A^{p-1}(X) / A^{p-1}(X)_\Z) \times H^p(X, \Z) \to H^p(X, \SDC{p}).
\end{equation}
\end{lem}

\begin{proof} 
Since $X$ admits a finite good cover (\cite{B-T}), $H^p(X, \Z)$ is finitely generated. Now the lemma follows from the fact that $A^{p-1}(X) / A^{p-1}(X)_\Z$ is divisible. (An abelian group $A$ is said to be divisible, if the assignment $a \mapsto n a$ gives an epimorphism $A \to A$ for each $n \in \N$.) 
\end{proof}


\subsection{Cup product}

As in the case of ordinary cohomology groups, smooth Deligne cohomology groups also have an operation called the \textit{cup product}, due to Be\u{\i}linson (\cite{Be}). In \cite{Bry,E-V}, the cup product is induced from a product on the complexes
$$
\cup : \ \SDC{p} \otimes_\Z \SDC{q} \longrightarrow \SDC{p+q}.
$$
Using the resulting description on the level of \Cech cocycles, we introduce here the cup product on smooth Deligne cohomology.

\smallskip

Let $\U = \{ U_\alpha \}$ be an open cover of $X$. For cochains
\begin{gather*}
(k_{\alpha_0 \cdots \alpha_p}, 
\tau^0_{\alpha_0 \cdots \alpha_{p-1}}, \ldots, 
\tau^{p-1}_{\alpha_0}) \in \check{C}^p(\U, \SDC{p}), \\
(m_{\alpha_0 \cdots \alpha_q}, 
\theta^0_{\alpha_0 \cdots \alpha_{q-1}}, \ldots, 
\theta^{q-1}_{\alpha_0}) \in \check{C}^q(\U, \SDC{q}),
\end{gather*}
we define $(n, \omega^0, \ldots, \omega^{p+q-1}) \in \check{C}^{p+q}(\U, \SDC{p+q})$ by
\begin{align*}
n_{\alpha_0 \cdots \alpha_{p+q}} 
&= 
k_{\alpha_0 \cdots \alpha_p} m_{\alpha_p \cdots \alpha_{p+q}}, \\
\omega^0_{\alpha_0 \cdots \alpha_{p+q-1}}
&= 
k_{\alpha_0 \cdots \alpha_p} \theta^0_{\alpha_p \cdots \alpha_{p+q-1}}, \\
& \ \vdots \\
\omega^{q-1}_{\alpha_0 \cdots \alpha_p}
&= 
k_{\alpha_0 \cdots \alpha_p} \theta^{q-1}_{\alpha_p}, \\
\omega^q_{\alpha_0 \cdots \alpha_{p-1}}
&= 
\tau^0_{\alpha_0 \cdots \alpha_{p-1}} \wedge d \theta^{q-1}_{\alpha_{p-1}}, \\
& \ \vdots \\
\omega^{p+q-1}_{\alpha_0}
&= 
\tau^{p-1}_{\alpha_0} \wedge d \theta^{q-1}_{\alpha_0}.
\end{align*}
If we put $(k, \tau^0, \ldots, \tau^{p-1}) \cup (m, \theta^0, \ldots, \tau^{q-1}) = (n, \omega^0, \ldots, \omega^{p+q-1})$, then we have a homomorphism
\begin{equation}
\cup : \
\check{C}^p(\U, \SDC{p}) \otimes_\Z \check{C}^q(\U, \SDC{q}) 
\longrightarrow
\check{C}^{p+q}(\U, \SDC{p+q}). \label{formula:cup_product_cochain}
\end{equation}

\begin{dfn}
We define the \textit{cup product} on smooth Deligne cohomology to be the homomorphism 
$$
\cup : \
H^p(X, \SDC{p}) \otimes_\Z H^q(X, \SDC{q}) \longrightarrow 
H^{p+q}(X, \SDC{p+q})
$$
induced from (\ref{formula:cup_product_cochain}).
\end{dfn}

The cup product is independent of the choice of $\U$, and gives rise to a natural homomorphism. It is known that the cup product is associative: $(f_p \cup f_q) \cup f_r = f_p \cup (f_q \cup f_r)$, and is (graded) commutative: $f_p \cup f_q = (-1)^{pq} f_q \cup f_p$, where $f_i \in H^i(X, \SDC{i})$ for $i = p, q, r$. We note that the cup product on smooth Deligne cohomology groups is compatible with the wedge product of differential forms and with the cup product on ordinary cohomology groups, through the homomorphisms in Proposition \ref{prop:Deligne_coh_exact_seq}.


\subsection{Integration}
\label{subsec:integration}

Integration formulae for smooth Deligne cohomology by using \Cech cocycles were first described by Gaw\c{e}dzki \cite{Gaw} for low dimensional manifolds. The integration formula in this subsection is derived from a general result in \cite{G-T}.

\medskip

Let $X$ be a compact oriented smooth $d$-dimensional manifold possibly with boundary. Then $X$ admits a triangulation (see \cite{Mil-S,Mun}). For an open cover $\U = \{ U_\alpha \}_{\alpha \in \Aa}$ of $X$, we consider a pair $(K, \phi)$ consisting of a triangulation $K$ of $X$ and a map $\phi : K \to \Aa$ such that $\sigma \subset U_{\phi_\sigma}$ for all $\sigma \in K$, where $\phi_\sigma \in \Aa$ is the image of $\sigma$ under $\phi$. 

For $i = 0, \ldots, d$ we define the set of \textit{flags} of simplices $F_K(i)$ by
$$
F_K(i) = \{ \vec{\sigma} = (\sigma^{d-i}, \ldots, \sigma^d) |\ 
\sigma^p \in K, \ \dim \sigma^p = p, \
\sigma^{d-i} \subset \cdots \subset \sigma^d \}.
$$

For a pair $(K, \phi)$ and $(n, \omega^0, \ldots, \omega^d) \in \check{C}^{d+1}(\U, \SDC{d+1})$, we put
\begin{eqnarray*}
\int_{(K, \phi)} (n, \omega^0, \ldots, \omega^d) = 
\sum_{i=0}^d 
\sum_{\vec{\sigma} \in F_K(i)}
\int_{\sigma^{d-i}} 
\omega^{d-i}_{  \phi_{\sigma^d} \cdots \phi_{\sigma^{d-i}}  },
\end{eqnarray*}
where the integration $\int_{\sigma^{d-i}}$ uses the orientation of $\sigma^{d-i}$ induced by that of $X$ along the flag $\sigma^{d-i} \subset \cdots \subset \sigma^d$. The following lemma is derived from \cite{G-T}:

\begin{lem} \label{lem:integration_formulae_cochain}
Let $X$ be a compact oriented smooth $d$-dimensional manifold.  

(a)
If $\check{f} = (m, \omega^0, \ldots, \omega^d) \in \check{Z}^{d+1}(\U, \SDC{d+1})$ is a cocycle, then we have 
$$
\int_{(K_1, \phi^1)} \!\! \check{f}
- \int_{(K_0, \phi^0)} \!\! \check{f} 
=
\sum_{\vec{\sigma} \in F_K(d)}
\sum_{n = 0}^d
(-1)^{n}
m_{
\psi^0_{\sigma^d} \cdots \psi^0_{\sigma^n} 
\psi^1_{\sigma^n} \cdots \psi^1_{\sigma^0}
}.
$$
In the formula above, $K$ is a common subdivision of $K_0$ and $K_1$, and we defined $\psi^i : K \to \Aa$ by $\psi^i_\sigma = \phi^i \circ \iota_i(\sigma)$, where $\iota_i(\sigma) \in K_i$ is the simplex of the smallest dimension such that $\sigma \subset \iota_i(\sigma)$.

(b)
For $\check{g} = (k, \tau^0, \ldots, \tau^{d-1}) \in \check{C}^d(\U, \SDC{d+1})$ we have 
$$
\int_{(K, \phi)} \!\! D \check{g}
-
\int_{(\d K, \d \phi)} \!\! \check{g}
=
(-1)^{d} \!
\sum_{\vec{\sigma} \in F_K(d)}
k_{  \phi_{\sigma^d} \cdots \phi_{\sigma^{0}}  },
$$
where $(\d K, \d \phi)$ is the pair obtained by restricting $(K, \phi)$ to the boundary.

\end{lem}

\begin{dfn} 
Let $X$ be a compact oriented smooth $d$-dimensional manifold without boundary.  We define a homomorphism
\begin{eqnarray*}
\int_X : H^{d+1}(X, \SDC{d+1}) \to \R/\Z
\end{eqnarray*}
as follows. First, we take a good cover $\U$ of $X$ and a pair $(K, \phi)$. Next, for a class $f \in H^{d+1}(X, \SDC{d+1}) \cong \check{H}^{d+1}(\U, \SDC{d+1})$, we take a representative $\check{f} \in \check{Z}^{d+1}(\U, \SDC{d+1})$. Finally, we put $\int_Xf = \int_{(K, \phi)}\check{f} \mod \Z$.
\end{dfn}

By Lemma \ref{lem:integration_formulae_cochain}, we can see that the integration is defined consistently and gives rise to a natural homomorphism.

\begin{lem} \label{lem:properties_integration}

(a) Let $X$ be a compact oriented smooth $d$-dimensional manifold without boundary. We take non-negative integers $p$ and $q$ such that $p + q = d$. For $\alpha \in A^p(X) / A^p(X)_\Z$ and $f \in H^q(X, \SDC{q})$, we have
$$
\int_X \iota(\alpha) \cup f = \int_X \alpha \wedge \delta(f) \mod \Z.
$$
In particular, we have $\int_X \iota(\alpha) = \int_X \alpha \mod \Z$ for $\alpha \in A^d(X)/A^d(X)_\Z$.

(b) Let $W$ be a compact oriented smooth $(d+1)$-dimensional manifold with its boundary $\d W$. For $f \in H^{d+1}(W, \SDC{d+1})$, we have
$$
\int_{\d W} f|_{\d W} = \int_W \delta(f) \mod \Z.
$$
\end{lem}

\begin{proof}
By the definition of the cup product, we have $\iota(\alpha) \cup f = \iota(\alpha \wedge \delta(f))$ in $H^{p+q+1}(M, \SDC{p+q+1})$. Hence we obtain (a) by the definition of the integration. To prove (b), we define a homomorphism $\varphi : \check{Z}^{d+1}(\U, \SDC{d+1}) \to \check{C}^{d+1}(\U, \SDC{d+2})$ by $\varphi(f) = f$. Then we have $D \varphi(f) = (0, \ldots, 0, \delta(f))$. An application of Lemma \ref{lem:integration_formulae_cochain} (b) completes the proof.
\end{proof}


\subsection{Geometric interpretation}

In this subsection, we \textit{formally} account for the two geometric interpretations of smooth Deligne cohomology groups mentioned in Section \ref{sec:introduction}.

\medskip

To begin with, we recall a geometric interpretation of the cohomology groups with coefficients in $\Z$.

\begin{prop}[see \cite{Bry,Bry-M1}] \label{prop:classification_coh}
Let $X$ be a smooth manifold.

(a) The group of isomorphism classes of principal $\T$-bundles over $X$ is isomorphic to $H^2(X, \Z)$.

(b) The group of isomorphism classes of abelian gerbes over $X$ is isomorphic to $H^3(X, \Z)$.

(c) The group of isomorphism classes of abelian 2-gerbes over $X$ is isomorphic to $H^4(X, \Z)$.
\end{prop}

\begin{rem}
Theory of gerbes was introduced by Giraud \cite{Gi} originally in a context of non-abelian cohomology. However, in this paper, we restrict ourselves to an abelian case: the abelian gerbes in Proposition \ref{prop:classification_coh} (b) mean \textit{gerbes bound by $\un{\T}$} (\cite{Bry,Bry-M1,Gi}), or their equivalent constructs (\cite{H,Mur}). Similarly, the abelian 2-gerbes in Proposition \ref{prop:classification_coh} (c) mean \textit{2-gerbes bound by $\un{\T}$} (\cite{Bre,Bry-M2}).
\end{rem}

We can identify the space of connections on a principal $\T$-bundle over $X$ with the space of 1-forms $A^1(X)$. It is known that abelian gerbes (and equivalent constructs) admit a notion of connections \cite{Bry,H,Mur}, and we can identify the space of connections on an abelian gerbe over $X$ with the space of 2-forms $A^2(X)$. When we take these connections into account, the following classification is possible:

\begin{prop}[\cite{Bry,Bry-M1}] \label{prop:classification_Deligne_coh}
Let $X$ be a smooth manifold.

(a) 
The group of isomorphism classes of $\T$-bundles with connection over $X$ is isomorphic to $H^2(X, \SDC{2})$.

(b) 
The group of isomorphism classes of abelian gerbes with connection over $X$ is isomorphic to $H^3(X, \SDC{3})$.
\end{prop}

\begin{rem}
Strictly, there are two types of connections on abelian gerbes: \textit{connective structure} and \textit{curving} \cite{Bry}. However, the lower one (connective structure) does not matter in the classification. So, as connections on abelian gerbes, we talk about curvings only.
\end{rem}

So far, we reviewed rigorous facts. In an attempt to generalize these facts, the notion of ``higher abelian gerbes'' arises. (See \cite{C-M-W,D-K} for example.) Translating Proposition \ref{prop:classification_coh} word by word, we have a formal claim:
\begin{quote}
``The group of isomorphism classes of abelian $n$-gerbes over $X$ is isomorphic to $H^{n+2}(X, \Z)$.''
\end{quote}
This naive understanding of abelian $n$-gerbes suffices for this paper. In a similar way, we obtain one of the geometric interpretations of smooth Deligne cohomology groups as the following formal claim:
\begin{quote}
``The group of isomorphism classes of abelian $n$-gerbes with connection over $X$ is isomorphic to $H^{n+2}(X, \SDC{n+2})$.''
\end{quote}

\smallskip

Granting these claims, we can also find geometric meaning of some related cohomology groups and homomorphisms: we can think of the homomorphism $\delta : H^{n+2}(X, \SDC{n+2}) \to A^{n+2}(X)_\Z$ as the operation of taking the ``curvature'' of a connection on an abelian $n$-gerbe. The $H^{n+1}(X, \R/\Z)$ classifies the ``flat abelian $n$-gerbes,'' and $H^{n+1}(X, \R) / H^{n+1}(X, \Z)$ the ``flat connections'' on a fixed abelian $n$-gerbe. The integration in Subsection \ref{subsec:integration} admits the interpretation as taking the ``holonomy'' of a connection on an abelian $n$-gerbe around the $(n+1)$-dimensional manifold $X$. 

\bigskip

Now we deduce the other geometric interpretation of smooth Deligne cohomology groups from the geometric interpretation above. Consider the following exact sequence of groups obtained by combining the homomorphisms in Proposition \ref{prop:Deligne_coh_exact_seq} (b):
$$
H^{n+1}(X, \SDC{n+1}) \overset{\delta}{\to}
A^{n+1}(X) \to
H^{n+2}(X, \SDC{n+2}) \overset{\chi}{\to}
H^{n+2}(X, \Z).
$$
A plausible claim would be that ``the space of connections on an abelian $n$-gerbe over $X$ is identified with $A^{n+1}(X)$.'' Then the exactness at the second term suggests that $H^{n+1}(X, \SDC{n+1})$ acts on the space of connections on an abelian $n$-gerbe leaving the isomorphism class of the underlying abelian $n$-gerbe fixed. So we can interpret the smooth Deligne cohomology group $H^{n+1}(X, \SDC{n+1})$ as the ``gauge transformation group of an abelian $n$-gerbe.'' 

\smallskip

In the case of $n = 0$, there is a natural isomorphism $H^1(X, \SDC{1}) \cong C^\infty(X, \T)$. This is nothing but the gauge transformation group of a principal $\T$-bundle (abelian 0-gerbe) over $X$, and the interpretation is justified. In the case of $n = 1$, we can also justify the interpretation by taking a particular geometric model of an abelian gerbe, such as in \cite{Bry,H,Mu-S}. Whatever model we take, we find that a principal $\T$-bundle gives an ``automorphism of an abelian gerbe,'' and that a principal $\T$-bundle with connection transforms connections on the abelian gerbe. Taking the isomorphism classes of principal $\T$-bundles with connection, we recover the smooth Deligne cohomology group $H^2(X, \SDC{2})$ by Proposition \ref{prop:classification_Deligne_coh} (a).


\section{Central extensions}
\label{sec:central_ext}

In this section, we construct central extensions of smooth Deligne cohomology groups, and prove Theorem \ref{ithm:central_ext} in Section \ref{sec:introduction} as Theorem \ref{thm:non_trivial} and \ref{thm:trivial}.

Throughout this section, $n$ denotes a fixed non-negative integer, $M$ a compact oriented smooth $(2n+1)$-dimensional manifold without boundary, and $\G(M)$ the smooth Deligne cohomology group $\G(M) = H^{n+1}(M, \SDC{n+1})$.


\subsection{Construction of central extensions}

\begin{dfn}
We define a group 2-cochain $S_M : \G(M) \times \G(M) \to \R/\Z$ by $S_M(f, g) = \int_M f \cup g$, where $\cup$ and $\int_M$ are the cup product and the integration for smooth Deligne cohomology, respectively.
\end{dfn}

\begin{lem}
The group 2-cochain $S_M$ is a cocycle, i.e.\@ we have
$$
S_M(f, g) + S_M(f + g, h) = S_M(f, g + h) + S_M(g, h)
$$
for $f, g, h \in \G(M)$.
\end{lem}

\begin{proof}
This lemma is clear because the cup product and the integration for smooth Deligne cohomology are homomorphisms.
\end{proof}

\begin{dfn}
We define a group $\til{\G}(M)$ by the set $\G(M) \times \T$ endowed with the group multiplication given by $(f, u) \cdot (g, v) = (f + g, uv \exp2\pi\im S_M(f, g))$.
\end{dfn}

By definition, the group $\til{\G}(M)$ is a central extension of $\G(M)$ by $\T$:
\begin{equation}
\begin{CD}
1 @>>> \T @>>> \til{\G}(M) @>>> \G(M) @>>> 1,
\end{CD}
\label{formula:central_ext_group}
\end{equation}
where homomorphisms $\T \to \til{\G}(M)$ and $\til{\G}(M) \to \G(M)$ are given by $u \mapsto (0, u)$ and $(f, u) \mapsto f$, respectively.

\begin{lem}
The commutator of $(f, u), (g, v) \in \til{\G}(M)$ is
\begin{equation}
\begin{split}
[(f, u), (g, v)] 
&= (0, \exp2\pi\im \left\{ S_M(f, g) - S_M(g, f) \right\}) \\
&= (0, \exp2\pi\im (1 + (-1)^n) S_M(f, g)).
\end{split} \label{formula:commutator}
\end{equation}
\end{lem}

\begin{proof}
We can verify the first line in the formula using that the inverse element of $(f, u) \in \til{\G}(M)$ is $(f, u)^{-1} = (-f, u^{-1}\exp2\pi\im S_M(f, f))$. The second line follows from the graded commutativity of the cup product.
\end{proof}


\subsection{Non-triviality in the case of $n$ even}

\begin{thm} \label{thm:non_trivial}
If $n$ is even, then $\til{\G}(M)$ is non-trivial.
\end{thm}

\begin{proof}
Clearly, a trivial central extension of an abelian group is abelian. Thus, it suffices to show that $\til{\G}(M)$ is a non-abelian group. By (\ref{formula:commutator}), the group $\til{\G}(M)$ is non-abelian if and only if there are elements $f, g \in \G(M)$ such that $2S_M(f, g) \neq 0$ in $\R/\Z$. Let $\alpha \in A^n(M)$ be an $n$-form on $M$ such that $d\alpha \neq 0$. If we take a Riemannian metric $g$ on $M$, then there is an $(n+1)$-form $\beta$ on $M$ such that $\alpha = d^*\beta = - *{\!}d{\!}*{\!}\beta$ by the Hodge decomposition theorem \cite{Wa}. Note that
$$
S_M(\alpha, -{\!}*{\!}\beta) 
= \int_M g(\alpha, \alpha) \dvol_g \mod \Z.
$$
We can take $r \in \R$ so that $2r^2 \int_M g(\alpha, \alpha) \dvol_g \not \in \Z$. Thus we have $2S_M(f, g) \neq 0$ taking $f, g$ as $r \alpha, -r *{\!}\beta \in A^n(M)/A^n(M)_\Z \subset \G(M)$, respectively. 
\end{proof}

If $n = 0$ and $M = S^1 = \T$, then $\G(S^1)$ is naturally isomorphic to the free loop group $L\T$. In this case, we can derive the non-triviality of $\til{\G}(S^1)$ from a result of Brylinski and McLaughlin \cite{Bry-M2}:

\begin{prop}
The central extension $\til{\G}(S^1)$ is isomorphic to $\h{L\T}/{\Z_2}$, where $\h{L\T}$ is the universal central extension of $L\T$.
\end{prop}

\begin{proof}
First of all, we describe the group 2-cocycle defining the universal central extension of $L\T$ given in \cite{P-S}: for a loop $f : S^1 \to \T$, we can find a map $F : \R \to \R$ such that $f = e^{2\pi\im F}$. We put $\Delta_F = F(1) - F(0)$ and $\overline{F} = \int_0^1F(t)dt$. The group 2-cocycle $c : L\T \times L\T \to \R/\Z$ defining $\h{L\T}$ is
$$
c(f, g) = \frac{1}{2}
\left\{
\int_0^1 F(t) \frac{dG}{dt}(t) dt  
+ \overline{F}\Delta_G + \Delta_F(\overline{G} - G(0))
\right\},
$$
and that defining $\h{L\T}/\Z_2$ is $2c$. Next, we express $S_M(f, g)$ in terms of $F$ and $G$. By taking a good cover and a triangulation of $S^1$, we perform a computation along the definition to obtain the following formula given in \cite{Bry-M2}:

$$
S_M(f, g) = 
\int_0^1 F(t) \frac{dG}{dt}(t) dt
- \Delta_F G(0).
$$
Finally, we introduce a group 1-cochain $T : L\T \to \R/\Z$ by $T(f) = \Delta_F \overline{F}$. Then we obtain an equality of 2-cocycles $S_M - \delta T = 2 c$, where $(\delta T)(f, g) = T(g) - T(fg) + T(f)$. Hence the central extensions determined by $S_M$ and $2 c$ are isomorphic.
\end{proof}

\begin{rem}
An interesting problem would be whether we can construct in a uniform way a central extension $\widehat{\G}(M)$ of $\G(M)$ such that $\widehat{\G}(M)/\Z_2 \cong \til{\G}(M)$ for $n = 2k > 0$. Such a central extension would play roles similar to those of the universal central extension $\h{L\T}$.
\end{rem}


\subsection{Triviality in the case of $n$ odd}

In this subsection, we suppose that $n = 2k + 1$. 

Recall the isomorphism of abelian groups given in Lemma \ref{lem:splitting_Deligne}:
$$
(A^{2k+1}(M) / A^{2k+1}(M)_\Z) \times H^{2k+2}(M, \Z) \cong \G(M).
$$
Taking the splitting $\sigma : H^{2k+2}(M, \Z) \to \G(M)$ inducing the isomorphism above, we obtain the following expression of the 2-cocycle $S_M$ by Lemma \ref{lem:properties_integration}:
$$
S_M((\alpha, c), (\alpha, c')) =
\int_M \alpha \wedge d\alpha' 
+ \int_M 
\{ \alpha \wedge \delta(\sigma(c')) + \alpha' \wedge \delta(\sigma(c)) \}
+ \int_M \sigma(c) \cup \sigma(c').
$$

\begin{lem}
There is a map $\tau : H^{2k+2}(M, \Z) \to \R/\Z$ such that 
$$
\int_M \sigma(c) \cup \sigma(c') = \tau(c + c') - \tau(c) - \tau(c').
$$
\end{lem}

\begin{proof}
We define an abelian group $E$ to be the set $H^{2k+2}(M, \Z) \times (\R/\Z)$ endowed with the group multiplication $(c, r) + (c', r') = (c + c', r + r' + \int_M \sigma(c) \cup \sigma(c'))$. Then $E$ is an extension of $H^{2k+2}(M, \Z)$ by $\R/\Z$:
$$
\begin{CD}
0 @>>> \R/\Z @>>> E @>>> H^{2k+2}(M, \Z) @>>> 0.
\end{CD}
$$
Because $H^{2k+2}(M, \Z)$ is finitely generated and $\R/\Z$ is divisible, the exact sequence above admits a splitting. Such splittings $s : H^{2k+2}(M, \Z) \to E$ correspond bijectively to the maps $\tau$ in the present lemma via $s(c) = (c, \tau(c))$.
\end{proof}

\begin{thm} \label{thm:trivial}
If $n$ is odd, then $\til{\G}(M)$ is trivial.
\end{thm}

\begin{proof}
Using the splitting $\sigma$ and the map $\tau$, we define $T : \G(M) \to \R/\Z$ by 
$$
T(\alpha, c) = 
\frac{1}{2}\int_M \alpha \wedge d\alpha 
+ \int_M \alpha \wedge \delta(\sigma(c)) 
+ \tau(c) \mod \Z.
$$
Note that $\frac{1}{2}\int_M \alpha \wedge d\alpha \in \R/\Z$ is well-defined for $\alpha \in A^{2k+1}(M)/A^{2k+1}(M)_\Z$ because of the Stokes theorem. A straightforward calculation shows $S_M(f, g) = T(f+g) - T(f) - T(g)$, which implies that $\til{\G}(M)$ is trivial.
\end{proof}

\begin{rem}
By the help of a result of Hopkins and Singer \cite{H-S} (Corollary 2.18, p.\ 17), we can also take a map $T : \G(M) \to \R/\Z$ such that $S_M(f, g) = T(f + g) - T(f) - T(g)$, provided some conditions on the $(4k+3)$-manifold $M$.
\end{rem}


\section{Reciprocity law}
\label{sec:reciprocity_law}

In this section, we prove our analogue of the Segal-Witten reciprocity law for the central extension $\til{\G}(M)$. 


\subsection{Complexification}

\begin{dfn}
Let $X$ be a smooth manifold. For a non-negative integer $p$, we define a complex of sheaves $\CSDC{p}$ by
$$
\CSDC{p} :\ 
\Z \longrightarrow
\un{A}^0_\C \overset{d}{\longrightarrow}
\un{A}^1_\C \overset{d}{\longrightarrow}
\cdots \overset{d}{\longrightarrow}
\un{A}^{p-1}_\C \longrightarrow
0 \longrightarrow \cdots,
$$
where $\un{A}^q_\C$ is the sheaf of germs of $\C$-valued differential $q$-forms.
\end{dfn}

The hypercohomology group $H^p(X, \CSDC{p})$ fits into exact sequences similar to (\ref{exact_seq:SDC_forms}) and (\ref{exact_seq:SDC_cohomology}) in Proposition \ref{prop:Deligne_coh_exact_seq}:
\begin{gather}
0 \to
H^{p-1}(X, \C/\Z) \to
H^p(X, \CSDC{p}) \overset{\delta}{\to}
A^p(X, \C)_{\Z} \to 0, 
\label{exact_seq:CSDC_forms} \\
0 \to
A^{p-1}(X, \C) / A^{p-1}(X, \C)_{\Z} \overset{\iota}{\to}
H^p(X, \CSDC{p}) \overset{\chi}{\to}
H^p(X, \Z) \to 0,
\label{exact_seq:CSDS_cohomology}
\end{gather}
where $A^q(X, \C)$ denotes the group of $\C$-valued $q$-forms on $X$, and $A^q(X, \C)_{\Z}$ the subgroup of closed integral $q$-forms. The formula for the cup product on $H^p(X, \SDC{p})$ induces
$$
\cup : H^p(X, \CSDC{p}) \otimes_\Z H^q(X, \CSDC{q}) \to
H^{p+q}(X, \CSDC{p+q}).
$$
Similarly, the integration formula for $H^p(X, \SDC{p})$ induces
$$
\int_X : H^{d+1}(X, \CSDC{d+1}) \to \C/\Z,
$$
where $X$ is supposed to be a compact oriented smooth $d$-dimensional manifold without boundary.

\medskip

Now we put $\G(X)_\C = H^{n+1}(X, \CSDC{n+1})$ for a non-negative integer $n$ and a smooth manifold $X$. If $M$ is a compact oriented smooth $(2n+1)$-dimensional manifold without boundary, then we can define a group 2-cocycle $S_{M, \C} : \G(M)_\C \times \G(M)_\C \to \C/\Z$ by $S_{M, \C}(f, g) = \int_M f \cup g$. Accordingly, we obtain a central extension $\til{\G}(M)_\C$ of $\G(M)_\C$ by $\C^* = \C - \{ 0 \}$:
$$
\begin{CD}
1 @>>> \C^* @>>> \til{\G}(M)_\C @>>> \G(M)_\C @>>> 1.
\end{CD}
$$
If $n$ is even, then $\til{\G}(M)_\C$ is a non-trivial central extension as well.


\subsection{Chiral and anti-chiral subgroups}

Let $k$ be a non-negative integer, $W$ a compact oriented smooth $(4k+2)$-dimensional manifold possibly with boundary, and $g$ a Riemannian metric on $W$. The Hodge star operator $* : A^{2k+1}(W, \C) \to A^{2k+1}(W, \C)$ satisfies $** = -1$. Thus, we have the following eigenspace decomposition:
\begin{equation}
A^{2k+1}(W, \C) = A^{2k+1}(W, \C)^+ \oplus A^{2k+1}(W, \C)^-,
\label{decomposition:diff_forms}
\end{equation}
where $A^{2k+1}(W, \C)^\pm = \{ \omega \in A^{2k+1}(W, \C) |\ \omega \mp \im *\omega = 0 \}$. 

Notice that any Riemannian metric conformally equivalent to $g$ induces the identical Hodge star operator on $A^{2k+1}(W, \C)$, so that the decomposition (\ref{decomposition:diff_forms}) is invariant under conformal transformations.

\begin{dfn}
Let $k$ be a non-negative integer, and $W$ a compact oriented smooth $(4k+2)$-dimensional Riemannian manifold possibly with boundary. We define the \textit{chiral subgroup} $\G(W)_\C^+$ and the \textit{anti-chiral (achiral) subgroup} $\G(W)_\C^-$ in the cohomology group $\G(W)_\C = H^{2k+1}(W, \CSDC{2k+1})$ to be
$$ 
\G(W)^\pm_\C = 
\{ f \in \G(W)_\C |\ 
\delta(f) \mp \im * \delta(f) = 0\}.
$$
\end{dfn}

It is easy to see that $\G(W)_\C^+ \cap \G(W)_\C^- = H^{2k}(W, \C/\Z)$. Since the Hodge star operator on $A^{2k+1}(W, \C)$ is invariant under conformal transformations, so are the subgroups $\G(W)_\C^\pm$.

\smallskip

In the case of $k = 0$, a Riemannian metric on $W$ induces a complex structure on $W$. Hence $W$ is a Riemann surface. The group $\G(W)_\C$ is naturally identified with $C^\infty(W, \C/\Z)$. For $f : W \to \C/\Z$, we have 
\begin{align*}
\delta(f) - \im * \delta(f) &= 2 \bar{\d} f, &
\delta(f) + \im * \delta(f) &= 2 \d f.
\end{align*}
Thus, in this case, $\G(W)_\C^+$ is the group of holomorphic functions $f : W \to \C/\Z$, and $\G(W)_\C^-$ is the group of anti-holomorphic functions $f : W \to \C/\Z$.

\begin{rem}
We call a $2k$-form $\omega \in A^{2k}(W, \C)$ such that $d \omega \in A^{2k+1}(W, \C)^+$ a \textit{chiral $2k$-form} (see \cite{W}). A chiral $2k$-form induces an element belonging to the subgroup $\G(W)_\C^+$ via the monomorphism in (\ref{exact_seq:CSDS_cohomology}). This is the reason that the subgroup $\G(M)_\C^+$ is termed chiral.
\end{rem}


\subsection{Reciprocity law}

Let $n$ be a non-negative integer. We suppose that $n$ is even: $n = 2k$. Let $W$ be compact oriented smooth $(4k+2)$-dimensional manifold with boundary. Then there is a natural homomorphism $r : \G(W)_\C \to \G(\d W)_\C$ induced by the restriction to the boundary.

If we pull the non-trivial central extension $\til{\G}(\d W)_\C$ back to $\G(W)_\C$ by $r$, then we obtain a central extension $r^*\til{\G}(\d W)_\C$ of $\G(W)_\C$:
$$
\begin{CD}
1 @>>> \C^* @>>> \til{\G}(\d W)_\C @>>> \G(\d W)_\C @>>> 1 \\
@. @| @AAA @AA{r}A @. \\
1 @>>> \C^* @>>> r^*\til{\G}(\d W)_\C @>>> \G(W)_\C @>>> 1. 
\end{CD}
$$
Clearly, $r^*\til{\G}(\d W)_\C$ is identified with $\G(W)_\C \times \C^*$ as a set, and the group multiplication is given by $(f, u) \cdot (g, v) = (f + g, uv \exp2\pi\im (r^*S_{\d W, \C}(f, g)))$. By the help of Lemma \ref{lem:properties_integration}, the group 2-cocycle $r^*S_{W, \C} : \G(W)_\C \times \G(W)_\C \to \C/\Z$ has the following expression:
$$
r^*S_{\d W, \C}(f, g) = 
\int_W \delta(f) \wedge \delta(g) \mod \Z.
$$

\begin{dfn}
Taking a Riemannian metric on $W$, we define a map $E_W : \G(W)_\C \to \C$ by
$$
E_W(f) 
= \pi \int_W \delta(f) \wedge * \delta(f)
= 2\pi\im \int_W \frac{\delta(f) + \im * \delta(f)}{2} \wedge 
                 \frac{\delta(f) - \im * \delta(f)}{2}.
$$
\end{dfn}

We can think of the following lemma as the ``Polyakov-Wiegmann formula'' for the ``energy term'' $E_W(f)$.

\begin{lem} \label{lem:PW_formula}
For $f, g \in \G(W)_\C$ we have
$$
2\pi\im (r^*S_{\d W, \C}(f, g)) =
E_W(f + g) - E_W(f) - E_W(g) + \Gamma_W(f, g),
$$
where $\Gamma_W : \G(W)_\C \times \G(W)_\C \to \C$ is defined by
$$
\Gamma_W(f, g) = 
\pi\im \int_W 
\left( \delta(f) - \im * \delta(f) \right) \wedge 
\left( \delta(g) + \im * \delta(g) \right).
$$
\end{lem}

\begin{proof}
By direct calculations, we obtain
\begin{align*}
E(f + g) - E(f) - E(g) 
&=
\pi\int_W 
\left( \delta(f) \wedge * \delta(g) 
     - * \delta(f) \wedge \delta(g) \right), \\
2\pi\im (r^*S_{W, \C}(f, g)) 
&= 
\pi\im \int_W
\left( \delta(f) \wedge \delta(g) + *\delta(f) \wedge *\delta(g) \right).
\end{align*}
These formulae establish the lemma.
\end{proof}

\begin{thm}
Let $W$ be a compact oriented smooth $(4k+2)$-dimensional Riemannian manifold with boundary. Then the central extension $r^*\til{\G}(\d W)_\C$ splits over the subgroup $\G(W)^+_\C$.
$$
\xymatrix{
1 \ar[r] & 
\C^* \ar[r] & 
r^*\til{\G}(\d W)_\C \ar[r] & 
\G(W)_\C \ar[r] & 
1 \\
       &
       &
       & 
\G(W)^+_\C \ar@{-->}[lu]^{\Phi_W}  \ar@{^{(}->}[u] &
}
$$
\end{thm}

\begin{proof}
Let us define a section $\Phi_W : \G(W)_\C \to r^*\til{\G}(\d W)_\C = \G(W)_\C \times \C^*$ by setting $\Phi_W(f) = (f, \exp E_W(f))$. Then Lemma \ref{lem:PW_formula} leads to 
$$
\Phi_W(f) \cdot \Phi_W(g) =
e^{\Gamma_W(f, g)} \Phi_W(f + g),
$$
where the dot means the multiplication in $r^*\til{\G}(\d W)_\C$. By the definition of $\Gamma_W$, we have $\Gamma_W(f, g) = 0$ for $f \in \G(W)^+_\C$. Hence $\Phi_W : \G(W)^+_\C \to r^*\til{\G}(\d W)_\C $ gives rise to a splitting of the central extension.
\end{proof}

Note that $r^*\til{\G}(\d W)_\C$ also splits over the subgroup $\G(W)^-_\C$.

As is remarked in Section \ref{sec:introduction}, in the case where $k = 0$ and $W$ is a Riemann surface, the theorem above recovers the Segal-Witten reciprocity law for the central extension $\h{L\T}/{\Z_2}$. (See \cite{Bry-M1,Bry-M2}.)

\bigskip

\begin{acknowledgment}
I am grateful to T. Kohno and M. Furuta for valuable discussion and 
useful suggestions.
\end{acknowledgment}


\begin{flushleft}
Graduate school of Mathematical Sciences, University of Tokyo, \\
Komaba 3-8-1, Meguro-Ku, Tokyo, 153-8914 Japan. \\
e-mail: kgomi@ms.u-tokyo.ac.jp
\end{flushleft}

\end{document}